\documentclass[12pt]{article}%
\usepackage{amsmath}
\usepackage{amsfonts}
\pdfoutput=1
\usepackage{amssymb}

\begin{document}

\author{G. Steinbrecher, N. Pometescu \\Department of Physics, University of Craiova, \\Str. A. I. Cuza, No.13, 200585 - Craiova, Romania}
\title{Minimization algorithm in the simulation of the Wall Touching Kink Modes}
\date{}
\maketitle

\begin{abstract}
The discretized variational principle in the simulation of the Wall Touching
Kink Modes (WTKM) is reformulated in terms of independent variables and a
corresponding constrained minimization algorithm is elaborated. In a frame of
a general formalism is proposed an algorithm for constrained linear
minimization adapted to this class of problems. The FORTRAN programme that
realize the algorithm is described.

\end{abstract}

\section{Introduction}

The simulation of the currents in the tokamak wall was studied previously in
\cite{AtanasiuZaharov1}, by using the boundary element method for solving the
MHD equations in the thin wall approximation by using the triangular linear
conforming finite element method. But a new problem arise when on the internal
face of the tokamak wall is welded a conducting plate (for instance a
limiter). In general, the position of the limiter is not related to the
existing triangulation, and so, nonconforming finite elements appears by the
triangulation of the limiter \cite{2}: the edges and vertices of the triangles
on the outer circumference of the limiter, in the generic case, are inside the
triangles resulted from the finite element study of the tokamak before the
limiter welding. Consequently, the physical data attached to the finite
elements of the limiter are related by linear constraints to the data attached
to the finite elements from the tokamak wall.

A problem to be solved is how to include these new constraints such that the
modifications in the existing code to be minimal. A specific problem of the
boundary elements method is that at each iteration step, large scale quadratic
optimization must be performed, where the Hessian matrix is not sparse.

Next, the article is organized as follows: In the first section will be
introduced the notations in order to formulate the general formalism for the
problem we intend to solve. For that, the (discretized) variational principle
is reformulated in the terms of independent variables. The new objective
function will be introduced in Section 2 and described an efficient algorithm
for constrained linear minimization that is adapted to this class of problems.
In Section 3 will be described the FORTRAN\ programme that realize the
algorithm and in the final section we give the conclusions.

\section{Notations}

We denote by $\mathcal{V}$, the set of all discretized variables (the
potentials and the currents) attached to the set of all vertices. This is
represented by a vector with $|\mathcal{V}|$ \ components $\mathbf{X}%
=\{X_{1},...,X_{|\mathcal{V}|}\}$ . Here, in general, $|\mathcal{A}|$ \ denote
the cardinality of the set $\mathcal{A}$. In this stage there are no
restrictions (conformity, Neumann or Dirichlet boundary conditions) on the
variables $\mathbf{X}=\{X_{1},...,X_{|\mathcal{V}|}\}$. The general form of
the functional after discretization is

\begin{align}
Q(\mathbf{X})  &  =\frac{1}{2}\left\langle \mathbf{X},\mathbf{HX}\right\rangle
+\left\langle \mathbf{L,X}\right\rangle +C\label{1}\\
&  =\frac{1}{2}\sum\limits_{m,n}X_{m}H_{m,n}X_{n}+\sum\limits_{m}X_{m}L_{m}+C
\label{2}%
\end{align}

Here $\mathbf{H}$ is the positive definite Hessian matrix of the quadratic
form \ $Q(\mathbf{X})$, constructed from the mutual capacitances and
inductances (\textit{see Eqs. (5.6, 5.9) }from \cite{AtanasiuZaharov1}). Here
$\left\langle \mathbf{L,X}\right\rangle $\ and $C$ are, respective, linear and
constant terms included for the sake of generality required for the
possibility to test the programme.

The subset of $\mathcal{V}$ \ that consists of all variables that are on the
boundaries (the welding line where the limiter is fixed to the wall, the inner
edge of the limiter and the variables associated on the boundary of the holes
in the tokamak wall) are subjected to restrictions denoted by $\mathcal{B}$ .
The rest, the set of independent variables, will be denoted by $\mathcal{I}$.
\ So,%
\[
\mathcal{V}=\mathcal{B}\cup\mathcal{I\ \ \ \ },\ \ \ \ \mathcal{B}%
\cap\mathcal{I}=\varnothing\ \ \ \
\]

\subparagraph{\textit{Remark} \ }

\textit{In the set }$\mathcal{B}$\textit{\ of boundaries we include, as usual,
the lines that define the holes in the tokamak wall, but we also include the
line of welding where the new metal plate (possibly a limiter) is\ attached to
the tokamak wall. In this last case the triangulation on the tokamak wall
remains the same as before the welding. A separate triangulation of the new
plate is performed. The vertex points the triangle lying on of the inner edge
\ (that are not in contact with the wall) are included in the set
}$\mathcal{B}$\textit{. \ The vertex points of the triangles \ of the plate
triangulation, that are also on the tokamak wall are included in the set
}$\mathcal{B}$\textit{. \ For a general case, these new points, appeared after
welding, are in the interior of some triangle constructed before welding, so
they are no more independent variables, they are subjected to linear
constraints resulted from linear interpolation. \ Consequently, the values of
the potentials on these new set of points are expressed, by linear
interpolation, by the values of the potentials on the triangles from the
tokamak wall.}

The general form of the constraints on the boundaries are of the form
\begin{equation}
X_{b}=\sum\limits_{i\in\mathcal{I}}F_{b,i}~X_{i}+S_{b};~b\in\mathcal{B}
\label{3}%
\end{equation}
where the matrix\ \ $F_{b,i}$ and the (possible) source term vector
$S_{b\text{ }}$ encodes the boundary condition. \ We denote the affine
submanifold of $\mathbf{R}^{|\mathcal{V}|}$ given by the constraints
Eqs.(\ref{3}) with $\mathbf{Z}$ , its dimension is $|\mathcal{V}%
|-|\mathcal{B}|=|\mathcal{I}|$

Our first goal is to develop a formalism that despite is not optimal, from the
point of view of memory management, it is sufficiently compact such that the
corresponding Fortran program is easy to be verified with synthetic data. In
this end \textbf{we expand }\ the arrays $\mathbf{F,S}$\ , that in
Eq.(\ref{3}) \ has low dimension, $|\mathcal{B}|\times|\mathcal{I}|$ ,
respectively $|\mathcal{B}|$\ \ to larger, the extended array $\widetilde
{\mathbf{F}}$ with dimensions $|\mathcal{V}|\times|\mathcal{V}|$ ,
respectively the extended array of the sources $\widetilde{\mathbf{S}}$ of
dimension $|\mathcal{V}|$ as follows.
\begin{align}
\widetilde{F}_{b,b^{\prime}}  &  =0;~b\in\mathcal{B},~b^{\prime}\in
\mathcal{B}\label{4}\\
\widetilde{F}_{i,j}  &  =\delta_{i,j};~i\in\mathcal{I},~j\in\mathcal{I}%
\label{5}\\
\widetilde{F}_{i,b}  &  =0;~~i\in\mathcal{I},b\in\mathcal{B}\label{6}\\
\widetilde{F}_{b,i}  &  =F_{b,i}=unchanged;~~i\in\mathcal{I},b\in
\mathcal{B}\nonumber
\end{align}
The corresponding expansion of the vector $\mathbf{S}$ is similar%
\begin{align}
\widetilde{S}_{i}  &  =0;~i\in\mathcal{I}\label{7}\\
\widetilde{S}_{b}  &  =S_{b}=unchanged;~b\in\mathcal{B}\nonumber
\end{align}
With these conventions we introduce the parametrization of the \ submanifold
\ $\mathbf{Z}$\ by the vector $\mathbf{Y}=\{Y_{1},...Y_{|\mathcal{V}|}\}$ of
the restrictions Eq.(\ref{3}) as follows \
\begin{align}
\mathbf{X}  &  \mathbf{=\widetilde{\mathbf{F}}Y+}\widetilde{\mathbf{S}}\\
X_{n}  &  =\sum\limits_{m\in\mathcal{V}}\widetilde{F}_{n,m}~Y_{m}%
+\widetilde{S}_{n};~n\in\mathcal{V} \label{8}%
\end{align}
where $\mathbf{Y}$ is an arbitrary vector with $|\mathcal{V}|$ components. By
Eqs.(\ref{4}, \ref{5}, \ref{6}, \ref{8}) results that in the case when%
\begin{equation}
Y_{i}=0,i\in\mathcal{I} \label{8.1}%
\end{equation}
results that
\[
\widetilde{\mathbf{F}}\mathbf{Y}=0
\]
Consequently without loss of generality in the parametrization from
Eq.(\ref{8}) we impose the restriction
\begin{equation}
Y_{b}=0;~b\in\mathcal{B} \label{9}%
\end{equation}
This subspace of the variable $\mathbf{Y}$ , will be denoted by $\mathbf{U}$ ,
it\ has the dimension $|\mathcal{I}|$, like the subspace $\mathbf{Z}$
\ defined by Eqs.(\ref{3}).

\section{The new objective function}

Now, the minimization problem of the objective function $ Q(\mathbf{X}) $ from
Eqs. (\ref{1}), (\ref{2}) with restriction given by Eq.(\ref{3}), or equivalently

\[
\min_{\mathbf{X\in Z}}Q(\mathbf{X})
\]
by the representations Eqs.(\ref{8}), (\ref{9}) can be reformulated as%

\begin{equation}
\underset{\mathbf{X\in Z}}{\min}Q(\mathbf{X})=\ \underset{\mathbf{Y\in
U}^{\prime}}{\min}Q^{(new)}(\mathbf{Y}) \label{10}%
\end{equation}

where the new quadratic form $Q^{(new)}(\mathbf{Y})$ is given by the following
set of equations%
\begin{align}
Q^{(new)}(\mathbf{Y})  &  =\frac{1}{2}\left\langle \mathbf{Y}%
,\mathbf{H^{(new)}Y}\right\rangle +\left\langle \mathbf{L^{(new)}%
,Y}\right\rangle +C^{(new)}=\nonumber\\
&  \frac{1}{2}\sum\limits_{m,n}Y_{m}H_{m,n}^{(new)}Y_{n}+\sum\limits_{m}%
Y_{m}L_{m}^{(new)}+C^{(new)} \label{11}%
\end{align}
According to Eqs. (\ref{1}, \ref{2} , \ref{8}) \ the new Hessian matrix is
given by%
\begin{align}
\mathbf{H}^{(new)}  &  =\widetilde{\mathbf{F}}^{T}\mathbf{H}\widetilde
{\mathbf{F}}\nonumber\\
H_{m,n}^{(new)}  &  =\sum\limits_{p.q}\widetilde{F}_{p,m}H_{p,q}\widetilde
{F}_{qn} \label{12}%
\end{align}
Similarly the new linear term is
\begin{align}
\mathbf{L^{(new)}}  &  \mathbf{=}\widetilde{\mathbf{F}}^{T}\mathbf{L+}%
\widetilde{\mathbf{S}}^{T}\mathbf{H}\widetilde{\mathbf{F}}\nonumber\\
L_{n}^{(new)}  &  =\sum\limits_{p.}\widetilde{F}_{p,n}L_{p}+\sum
\limits_{p.q}{}\widetilde{S}_{p}H_{p,q}\widetilde{F}_{qn} \label{13}%
\end{align}
By the same reasoning, the new constant term is%
\begin{align}
C^{(new)}  &  =C+\left\langle \mathbf{L,}\widetilde{\mathbf{S}}\right\rangle
+\frac{1}{2}\left\langle \widetilde{\mathbf{S}},\mathbf{H}\widetilde
{\mathbf{S}}\right\rangle \label{14}\\
C^{(new)}  &  =C+\sum\limits_{p.}\widetilde{S}_{p}L_{p}+\frac{1}{2}%
\sum\limits_{m,n}\widetilde{S}_{m}H_{m,n}\widetilde{S}_{n} \label{15}%
\end{align}

\section{The structure of the Fortran90 programmes.}

The programmes are written such that they can be used for a large class of
quadratic minimization problems.

\subsection{Generation of the initial data, without boundary conditions}

\subsubsection{ \ \ \ \ \ \ The Hessian matrices used in test}

The synthetic data for test must be chosen such that the quadratic form
associated to the Hessian matrix is positive definite, and the asymptotic
behavior for large indices must be similar to that of mutual capacities and
mutual inductance matrices from Ref. \cite{AtanasiuZaharov1} . We used two
forms%
\begin{equation}
H^{(1)}(m,n):=d\delta_{m,n}+\frac{1}{(m+n+a)^{p}};d>0;~~a>0;~p\in\mathbb{N}
\label{16}%
\end{equation}

respectively%

\begin{equation}
H^{(1)}(m,n):=d\delta_{m,n}+\left[  \frac{\sin[a(m-n)]}{(m-n)}\right]
^{p};d>0;~~a>0;~p\in\mathbb{N} \label{17}%
\end{equation}

It can be verified that these Hessian matrices are positive definite by
using the identities%

\begin{equation}
\frac{1}{(m+n+a)^{p}}   =\int\limits_{0}^{\infty}...\int\limits_{0}^{\infty
}dx_{1}..dx_{p}\exp\left[  -\left(  m+n+a\right)  \sum\limits_{k=1}^{p}%
x_{k}\right] 
\end{equation}

\begin{equation}
\left[  \frac{\sin[a(m-n)]}{(m-n)}\right]  ^{p}2^{p}  =\int\limits_{-a}%
^{a}...\int\limits_{a}^{a}dx_{1}..dx_{p}\exp\left[  i\left(  m-n\right)
\sum\limits_{k=1}^{p}x_{k}\right]
\end{equation}

\subsubsection{Programming details}

The initial data are generated such that the result of the constrained
optimization are already known. \textbf{The generation of the matrix }%
$H_{m,n}$ \textbf{and the array }$L_{m}$ \textbf{and constant }$C$ from
Eq.(\ref{2}) is performed in the module quadraticformdatamod. It has the
following entries:

\qquad\emph{module quadraticformdatamod}

\qquad\emph{implicit none \ }! contains all of the constant scalars, arrays,
matrices and their generating subroutines

\qquad\emph{integer, parameter:: nvariables=10}\textit{\ \ \ \ }! Number of
free variables in the objective function.

\textit{\qquad}\emph{real(8), parameter::hessa=0.0d0}\textit{\ \ \ \ }!
parameter in the test hessian function, shift , only for test runs

\textit{\qquad}\emph{real(8), parameter::hessdiag=1.0000d-4}\textit{\ \ }!
diagonal term of hessian, only for test runs

\textit{\qquad}\emph{integer, parameter::hessn=2}\textit{\ \ }! parameter in
the test hessian function, exponent , only for test runs

\textit{\qquad}\emph{real(8), dimension(:,:), ALLOCATABLE:: Hessian}%
\textit{\ \ \ }! Used in "objective function module", give the quadratic term
of objective\textit{\ }function

\textit{\qquad}\emph{real(8), dimension(:), ALLOCATABLE:: Linearterm}%
\textit{\ \ }! Used in "objective function module", give the linear term of
objective function

\textit{\qquad}\emph{real(8):: constantterm}\textit{\ \ }! Used in "objective
function module", give the constant term of objective function .

The initialization is controlled by the subroutine \ \emph{subroutine
initializQuadrform(errorflag)}

When called from the main program, activates the following subroutines:

\qquad\qquad\emph{subroutine allocatearrays(nvariables, succesfullallocated)}%
\textit{\ \ }

This subroutine allocate the Hessian matrix \ and the array of linear terms.
Their numerical values, as well as of the constant $C$ are fixed in the subroutines

\qquad\qquad\emph{subroutine generateHessianmatrix (nvariables, errorflagout)}

\textit{\qquad\qquad}\emph{subroutine generateLinearterm(nvariables,
errorflagout)}

\textit{\qquad\qquad}\emph{subroutine generateconstantterm(nvariables)}

For test runs the matrix elements of the Hessian matrix are provided by the function

\qquad\qquad\emph{function hessianfunct(nvariables, i, j, errorflaghfunct)
result(hess), having the heading:}

\textit{\qquad\qquad\qquad}\emph{integer, intent(in):: nvariables, i, j}

\textit{\qquad\qquad\qquad}\emph{integer, intent(out)::errorflaghfunct}

\ \ \ \ Its algebraic form is selected such that the resulting Hessian matrix
is pozitive. It contains free parameters defined in the front of this module:
\emph{parameter::hessa}, and \emph{parameter::hessdiag}.\newline The linear
term and constant term are generated such that the exact value of the
minimization is the result returned by the special choice of the following
real valued function:

\ \ \ \ \ \ \ \ \ \ \textit{\ \ }\emph{function lfunct(k)}\textit{\ .}

\subsection{Imposing boundary conditions.}

The generation of the matrix $F_{b,i}$ , the source term array is $S_{b}$ ,
from Eq.(\ref{3}), the generation of the new Hessian matrix $H_{m,n}^{(new)}$,
the new linear term $L_{m}^{(new)}$, the new constant term $C^{(new)}$, that
defines the new quadratic form from Eq.(\ref{11}) resulting from the
restriction of the quadratic form Eq.(\ref{2}) on the submanifold $\mathbf{Z}
$ imposed by the boundary conditions Eq.(\ref{3}), is realized according to
the equations (\ref{12}), (\ref{13}) and (\ref{15}). The explicit realization
is in the following module:

\emph{module boundarydatamod}\textit{\ }\newline It has the following entries

\emph{use quadraticformdatamod}

\emph{implicit none}

\qquad\emph{integer, parameter:: nboundaryelements= 40}\textit{\ ; }! Number
of variables to be eliminated by boundary conditions

\textit{\qquad}\emph{integer, dimension(:), ALLOCATABLE::boundarylist}

! boundarylist(k)=1 $=>$ variable $k$ is from set $\mathcal{B}$,
the boundary set, else is $=0$

\textit{\qquad}\emph{real(8), dimension(:), ALLOCATABLE:: Sbound }! encode
boundary sources, term $S(i)$ , zero for $i$ is in $\mathcal{I}$

\textit{\qquad}\emph{real(8), dimension(:,:), ALLOCATABLE:: Fboundarymatrix }!
here sparse matrix, encode boundary condition

\textit{\qquad}\emph{real(8), dimension(:,:), ALLOCATABLE:: newHessian}%
\textit{\ }! Used in "objective function module", give the quadratic term of
objective function

\textit{\qquad}\emph{real(8), dimension(:), ALLOCATABLE:: newLinearterm}%
\textit{\ }! Used in "objective function module", give the linear term of
objective function

\textit{\qquad}\emph{real(8):: newconstantterm}\textit{\ }! Used in "objective
function module", give the constant term of objective function

The allocation and generation of the new arrays is controlled by

\qquad\emph{subroutine initboundaryArrays(errorflag)}

\ \ \ \ \ \ It is activated from the main programme. \ By its call, finally
the matrix $F_{b,i}$ , the source term array is $S_{b}$ , from Eq.(\ref{3}%
)\ \ are allocated and computed. To this end, this subroutine activate the
following subroutines

\qquad\qquad A.

\qquad\qquad\emph{subroutine initboundarydata(errorflag)}

\qquad\qquad By its call, finally the matrix $F_{b,i}$ , the source term array
is $S_{b}$ , from Eq.(\ref{3})\ \ are allocated and computed. It controls the
following subroutines

\qquad\qquad\textit{A1 . }

\ \ \ \ \ \ \ \ \ \ \ \ \emph{subroutine allocateboundarydata(nvariables,
errorflag1)}

\qquad\qquad\textit{A2 \ }

\ \ \ \ \ \ \ \ \ \ \ \ \emph{subroutine generateboundarycond(nvariables,
errorflag2)}
\\

\qquad\qquad\qquad\textit{B. \ }

\qquad\qquad\qquad\emph{subroutine initializNewQuadrform(errorflag2)}

By calling this subroutine are allocated and computed the
new Hessian matrix $H_{m,n}^{(new)}$, the new linear term $L_{m}^{(new)}$, the
new constant term $C^{(new)}$. To this end the following subroutines are controlled:

\qquad\qquad\qquad\textit{B1.}

\qquad\qquad\qquad\emph{subroutine allocateNewarrays(nvariables,
succesfullallocated)}

\qquad\qquad\qquad\textit{B2.}

\qquad\qquad\qquad\emph{subroutine generateNewHessianmatrix (nvariables,
errorflhessiangen)}

\textit{\qquad\qquad\qquad B3.}

\qquad\qquad\qquad\emph{subroutine generateNewLinearterm(nvariables,
errorflgenlin)}

\qquad\qquad\qquad\textit{B4.}

\qquad\qquad\qquad\emph{subroutine generateNewconstantterm(nvariables)}%
\textit{\ \ }

By calling these previous subroutines the initialization phase of the
programme is finished.

The constrained optimization is encoded in the optimization programme that
used a slightly modified version of the Fletcher-Reeves conjugate gradient
method. Our new version is particulariation of the general nonlinear
optimization method to the case when the objective function is quadratic
polinomial. It returns exact result after a single optimization cycle, for an
ideal computer, or at most 2-3 iterations, due to rounding errors.

For the constrained optimization of the quadratic form defined in the
Eq.(\ref{11}), the algorithm uses the objective function, gradient and
Hessian of the objective function Eq.(\ref{11}). The Hessian is constant
and was already computed. The gradient and the objective functions are
 contained in the module

\qquad\qquad\emph{module Newobjective\_functionMod}\textit{\ }\ 

It has the entries

\qquad\qquad\emph{use quadraticformdatamod}

\qquad\qquad\emph{use boundarydatamod}

It contains the following realization of the objective function

\qquad\qquad\qquad\emph{function NewobjectivefunctionFunc(nvariables,
variables, errorflag) result(f)}

\qquad\qquad\qquad\emph{integer, intent(in):: nvariables}\textit{\ }! \# of parameters

\qquad\qquad\qquad\emph{real(8), intent(in)::variables(nvariables)}%
\textit{\ }\textbf{! variables}

\qquad\qquad\qquad\emph{integer,intent(out)::errorflag}

\qquad\qquad\qquad\emph{real(8)::f \ }! returned function value

The gradient is computed by the following subroutine

\qquad\qquad\qquad\textit{subroutine newgradSubr(nvariables, variables,
gradient, errorflag)}

\textit{\qquad\qquad\qquad integer, intent(in)::nvariables}

\textit{\qquad\qquad\qquad real(8), intent(in)::variables(nvariables)}

\textit{\qquad\qquad\qquad real(8), intent(out)::gradient(nvariables) }! The gradient

\textit{\qquad\qquad\qquad integer, intent(out)::errorflag}

In this module we have the subroutine, that Projects to the subspace denoted
by " U" defined by formula (\ref{11}).

\qquad\qquad\qquad\textit{subroutine projection(vector) \ }

\textit{\qquad\qquad\qquad real(8), intent(inout)::vector(nvariables) \ }

The constrained conjugate gradient optimization programme is contained in the \ module

\qquad\qquad\textit{module ProjFletcherReevesMod}

It has the first entries:

\qquad\qquad\qquad\textit{use quadraticformdataMod}

\textit{\qquad\qquad\qquad use boundarydataMod}

\textit{\qquad\qquad\qquad use newobjective\_functionMod}

It contains the constrained minimization programme

\qquad\qquad{\large \qquad}\textit{subroutine
ProjFletcherReevesSubr1(nvariable,nitmax, gradientbound, metric, x0, xf, pnit,
minvalue, gradientfinal, errorflag)}

The programme uses the following arguments

\qquad\qquad\qquad\textit{integer, intent(in)::nvariable }! nr of variables

\qquad\qquad\qquad\textit{real(8), intent(in):: nitmax }  ! max allowed nr
of iteration, stop criteria

\qquad\qquad\qquad\textit{real(8), intent(in)::gradientbound \qquad}! stop
criteria: if gradient module

\ gradientbound then stops

\qquad\qquad\qquad\textit{real(8), intent(in):: metric(nvariable) } !
for rescalling the variables

\qquad\qquad\qquad\textit{real(8), intent(in)::x0(nvariable) \qquad}! initial
point Side effect, it is modified

\qquad\qquad\qquad\textit{real(8), intent(out)::xf(nvariable) \qquad}! final point

\qquad\qquad\qquad\textit{real(8), intent(out):: pnit \qquad}! number of
actual iterations

\qquad\qquad\qquad\textit{real(8), intent(out)::minvalue \qquad}! The
final minimal value

\qquad\qquad\qquad\textit{real(8), intent(out)::gradientfinal \qquad}! module
of gradient value after optimization, if close to zero

\qquad\qquad\qquad\textit{integer, intent(out)::errorflag \qquad}! $=0$
\ optimization is succesfull, else $=1$ \textit{\ }

\textbf{The mainprogamme \ has the following entries}

\qquad\qquad\qquad\textit{program FRoptimizationmain}

\textit{\qquad\qquad\qquad use ProjFletcherReevesMod}

\textit{\qquad\qquad\qquad use quadraticformdatamod}

\textit{\ \ \ \ \qquad\qquad\ \ use boundarydatamod}

In this test programme we have succesively the following subroutine calls for initialization:

\qquad\qquad\qquad\qquad\textit{call initializQuadrform(errorflag)}

\qquad\qquad\qquad\qquad\textit{call initboundaryArrays(errorflag)}

The follwing call is for \ final test

\qquad\qquad\qquad\textit{call ProjFletcherReevesSubr1(nvariables,nitmax,
gradientbound, metric, variables1, variablesfin, pnit, minvalue,
gradientfinal, errorflag)}

If the programme is correct, the minimal value of the objective function must
be close to zero and the returned values of "variablesfin" must be close to
the selected already known exact solution.

\section{Conclusions}

In order to solve the problems that apear in the simulation of the WTKM, we
propose a new general algorithm for the construction of the new objective
function (that appears in the new optimization problem after attaching the
limiter on the tokamak wall). The construction of the new objective function
starts from the quadratic objective function, see \cite{AtanasiuZaharov1},
that appears in the simulation without limiter. The coupling of the currents
and electric potentials in the limiter and tokamak wall are described by a set
of linear constraints. By a suitable change of variables the initial
constraints are transformed, and the constrained optimization is greatly
simplified (compared to the general constrained conjugated gradient
optimization method \cite{3}). The FORTRAN90 test programme consists of main
programme and four modules, that

\qquad- allocate and generate the initial data, the Hessian matrix and linear
part of the objective function, as well as constant term, that is used for
verification. The synthetic data for second order term were chosen such that
the resulting matrix of relative capacitances of the triangulation be strictly
positive definite

\qquad- allocate and generate the data related to attaching the limiter, in
form of set of arrays, that defines the constraints related to boundary
conditions on the contact line between limiter and tokamak wall

\qquad- constructs the new Hessian and new linear term, that, generate the
objective function that describe the limiter-tokamak wall system.

\qquad- perform the constrained minimization.

The subroutines returns together to variables an error message, that in the
case of errors stop the execution.

An advantage of the conjugate gradient methods, in the Fletcher-Reeves version
\cite{4}, is that (at least when it is used for linear optimization) it can be
efficiently run on parallel computers, by computing the gradients and
conjugated directions of separate groups of variables on different processors.
This advantage persists also in the our version of the constrained
optimization.\medskip\medskip

\textbf{Acknowledgement} 

This work has been carried out within the framework
of the EUROfusion Consortium as a complementary project and has been received
funding from the Romanian National Education Minister/Institute of Atomic
Physics under contract 1EU-2/2//01.07.2016. The collaboration with Calin
Vlad Atanasiu is acknowledged.

\end{document}